# Piezoelectric Sensors for Real-time Monitoring and Quality Control in Additive Manufacturing

Rashid T. Momin

rashidmomin.2006@gmail.com

Department of Diploma in Mechanical Engineering (DME)

Veermata Jijabai Technological Institute (VJTI), Mumbai

Abstract:

Within the ever-evolving landscape of engineering, particularly in the dynamic domain of additive manufacturing, a pursuit of precision and excellence in production processes takes centre stage. This research paper serves to give a comprehensive understanding of piezoelectric sensors, a topic that is both academically engaging and of practical significance, catering to both seasoned experts and those newly venturing into the field. Additive manufacturing, lauded for its groundbreaking potential, underscores the imperative of rigorous quality control. This introduces the piezoelectric sensors, devices that may be unfamiliar to many but possesses considerable potential. This paper embarks on a methodical journey, commencing with an introductory elucidation of the piezoelectric effect. It then advances to the vital role of piezoelectric sensors in real-time monitoring and quality control, unveiling their potential and relevance for newcomers and seasoned professionals alike. This research, structured systematically from fundamental principles to pragmatic applications, presents findings that are not only academically informative but also represent a substantial stride towards achieving precision and high-quality manufacturing processes in the engineering field.

## I. Introduction

Piezoelectric sensors play a crucial role in real-time monitoring and quality control in the field of additive manufacturing. The development and application of these sensors have gained significant attention due to their ability to accurately measure mechanical parameters and provide insights into the performance and quality of manufactured parts. The findings aim to provide a comprehensive understanding of the piezoelectric effect and its combination with additive manufacturing technology. The piezoelectric effect refers to the phenomenon in which certain materials generate an electric charge when subjected to mechanical stress or pressure. This effect occurs due to the reorientation of electric dipoles within the material's structure, resulting in the conversion of mechanical energy into electrical energy. Piezoelectric materials, such as piezoelectric ceramics and polymers, have been widely utilized in the manufacture of intelligent devices including sensors, actuators, and transducers (Yang). These materials possess a unique property of coupling between their mechanical and electrical behaviours, making them ideal for sensing applications. Piezoelectric sensors in additive manufacturing offer various advantages over traditional sensors.

They exhibit flexibility, easy processing, and melding capabilities, making them suitable for integration into complex geometries during the additive manufacturing process. Additionally, piezoelectric sensors provide high sensitivity and accuracy in measuring mechanical parameters such as temperature, pressure, and vibration. Moreover, the combination of piezoelectric sensors with additive manufacturing technology enables the fabrication of customized and embedded sensors within the structure of manufactured parts, eliminating the need for additional sensors and minimizing the risk of damage or interference (Ding).

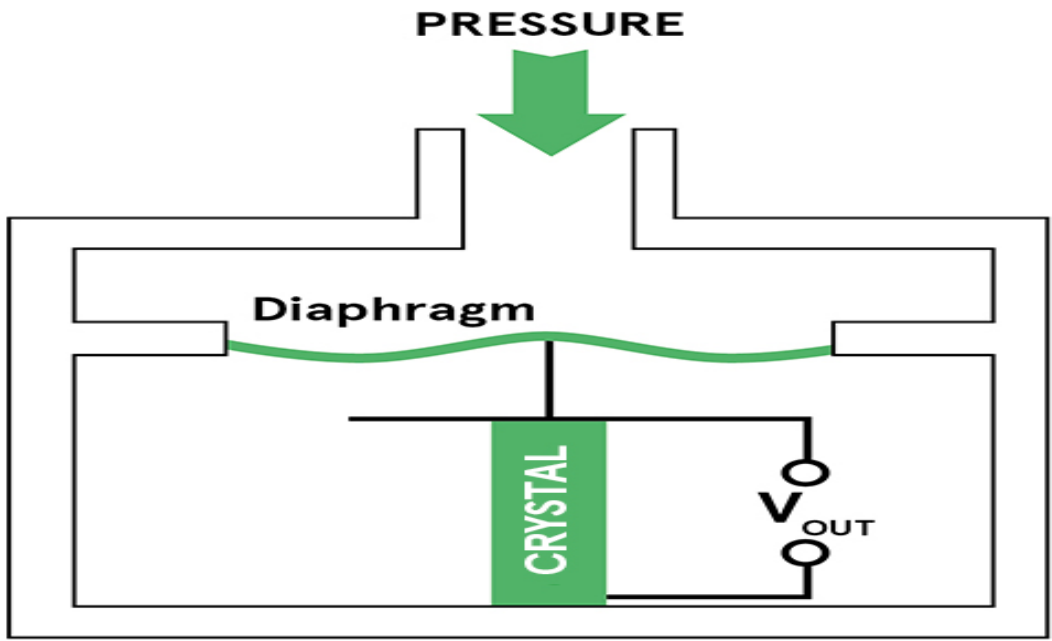

Fig 1, Simplified diagram of a Piezoelectric Pressure Sensor (Avnet Abacus)

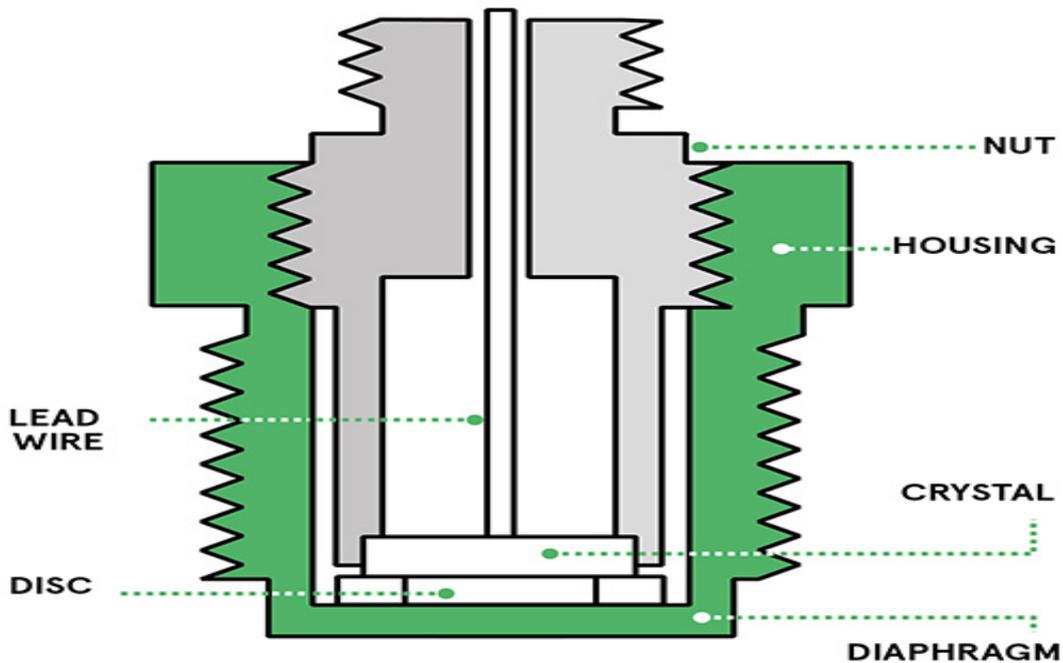

Fig 2, Cross-Section of a Piezoelectric Sensor Construction (Avnet Abacus)

One of the key challenges in additive manufacturing is ensuring the quality and precision of the manufactured parts. Defects, such as voids, porosity, and dimensional inaccuracies, can adversely affect the performance and reliability of the final product. Therefore, the integration of piezoelectric sensors in additive manufacturing allows for monitoring and quality control during the fabrication process. This can help identify and rectify any potential defects or variations in real-time, ensuring that the manufactured parts meet the required standards and specifications. This means that the sensors can be programmed to adjust and optimize the manufacturing process based on real-time feedback, improving efficiency, productivity, and overall product quality. Adaptive manufacturing using piezoelectric sensors in additive manufacturing can also enable the production of highly complex and intricate structures that are not feasible with conventional manufacturing techniques (Burgos)

By embedding piezoelectric sensors within the additive manufacturing process, it is possible to monitor and control the deposition of material layers in real-time. This ensures precise layer-by-layer fabrication, leading to improved dimensional accuracy and overall quality of the manufactured parts. Additionally, the integration of piezoelectric sensors in additive manufacturing enables the implementation of structural health monitoring systems (Ju). These systems can detect and assess the structural integrity of manufactured parts during their service life, providing early warning signs of potential failures or damages. Moreover, the combination of piezoelectric sensors and additive manufacturing allows for the development of smart structures that can adapt to changing environmental conditions. For example, employing 3D printing to prepare novel aggregate-shape embedded piezoelectric sensors. These sensors have the ability to sense and respond to external stimuli such as mechanical stress, temperature changes, and vibrations. By utilizing piezoelectric sensors in additive manufacturing, it becomes possible to create intelligent structures that can actively detect and respond to their surroundings (Ju).

The integration of piezoelectric sensors in additive manufacturing is particularly beneficial for sensing and monitoring the printing process itself. Real-time monitoring of the printing process using piezoelectric sensors allows for immediate detection and correction of any defects or inaccuracies.
This results in improved print quality and reduces the need for post-processing or rework. Furthermore, the use of piezoelectric sensors in additive manufacturing can lead to advancements in material development and optimization (Ding).

In addition to material optimization, additive manufacturing with piezoelectric sensors also allows for the production of functional composites. As we embark on this research journey, our goal is two-fold: to thoroughly understand piezoelectric sensors and to showcase their potential in Additive Manufacturing. With this foundational knowledge, we're prepared to dive deeper into the practical applications and implications of these sensors. In the upcoming sections, we'll explore the fusion of science and engineering, where piezoelectric sensors are poised to redefine quality assurance and precision in Additive Manufacturing.

## II. Definition and Glossary

1. Piezoelectric sensors: Sensors that generate an electric charge in response to applied mechanical stress or vibrations. They can be attached to the surface of a structure or embedded within a composite structure, making them highly versatile and suitable for complex configurations.

2. Additive manufacturing, also known as 3D printing, is a revolutionary technology that has transformed various industries, including sensor design and manufacturing. This technology offers new opportunities in the fabrication of piezoelectric sensors for structural health monitoring (source). By employing additive manufacturing techniques, such as 3D printing, piezoelectric materials, including piezoceramics, polymers, or composites, can be easily processed into sensor component.

3. Piezoelectric Effect The piezoelectric effect refers to the ability of certain materials to generate an electric charge when subjected to mechanical stress or vibrations. Piezoelectric sensors play a crucial role in additive manufacturing by enabling real-time monitoring of the printing process. The piezoelectric effect is a key principle in additive manufacturing, as it allows for the real-time monitoring of the printing process.

4. Structural Health Monitoring: The process of assessing and monitoring the condition of structures to ensure their safety, reliability, and longevity. Piezoelectric sensors integrated into additive manufacturing enable real-time monitoring of the structural health of printed components. By detecting changes in vibration, strain, and other mechanical properties, these sensors can provide valuable data on the structural integrity of the printed objects. SHM systems utilizing piezoelectric sensors can contribute to early detection of defects, preventing potential failures and enhancing overall safety.

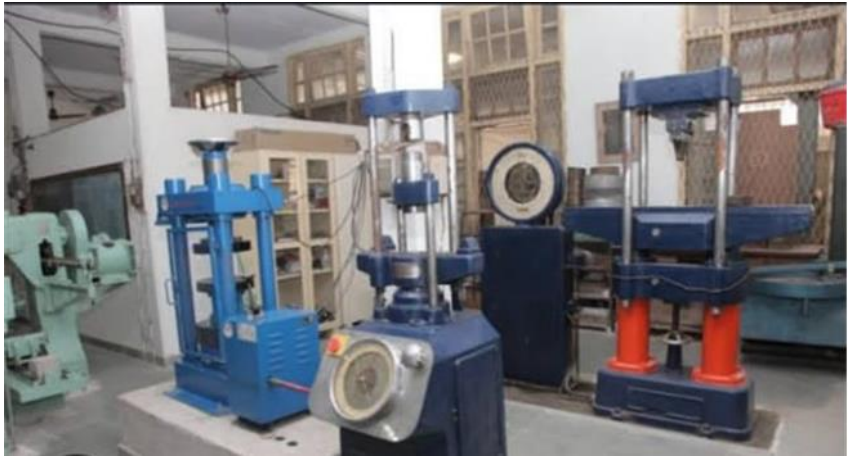

Fig 3, UTM (Universal Testing Machine) in Structural Engineering Dept., VJTI

5. Polyvinylidene Fluoride: A commonly used piezoelectric polymer material in additive manufacturing due to its favourable piezoelectric properties. In recent years, additive manufacturing has emerged as a potential approach for sensor design and manufacturing, including the fabrication of piezoelectric sensors. Using additive manufacturing technology, researchers have successfully incorporated piezoelectric materials, such as polyvinylidene fluoride, into sensor components for improved sensing capabilities (Tselikos).

## III. Literature Review

Manufacturing industries have witnessed significant progress in quality control methodologies in recent years. This literature review explores key research papers to elucidate the advancements in quality control and manufacturing processes.

Paper 1: "New Approach for Quality Control in Manufacturing Process" by Muhammad Radhi bin Hamzah et al.

This paper emphasizes and focuses on minimizing defects and ensuring product quality by employing a controlled strategy and a methodological approach. The use of ISO standards in Malaysia and the implementation of photoelectric sensors at Quality Control (QG) stations are presented as vital components to achieve zero defect outflow. This approach emphasizes the need for technical equipment and the integration of physical visual inspection techniques with modern technology. By ensuring product quality through systematic control and cutting-edge technology, the study underlines the evolution of quality standards in manufacturing. importance of minimizing defects in the manufacturing process using a controlled strategy and technology, such as photoelectric sensors at Quality Control (QG) stations. The primary focus is on defect reduction and quality control in manufacturing. The study emphasizes the need for technical equipment and integration of physical visual inspection techniques with modern technology. (Muhammad Radhi bin Hanzah)

Paper 2: "Research on the Application Status of Machine Vision Technology in Furniture Manufacturing Process" by Rongorongo Li et al.

This study examines the application of Machine Vision (MV) technology in furniture manufacturing. MV is utilized for personnel safety monitoring, equipment efficiency enhancement, data collection at various manufacturing stages, and quality assurance to prevent the release of defective items. Furthermore, MV enables automatic sorting of furniture parts, contributing to product classification and storage. The research highlights how MV transforms the production management landscape by enhancing automation and production efficiency. The paper anticipates a future integration of 3D MV with other advanced manufacturing technologies, creating intelligent and efficient manufacturing workshops. (Li)

Paper 3: "Design of a Plantar Pressure Insole Measuring System Based on Modular Photoelectric Pressure Sensor Unit" by Bin Ren and Jianwei Liu.

This research introduces a novel plantar pressure sensing insole based on photoelectric sensing technology. The modular pressure sensor is designed to offer simplicity, reliability, and cost-effectiveness, targeting applications in wearable human body assist equipment. The modular sensor unit is developed with a focus on sensing principles, structural design, and elastic materials. It enables calibration analysis and proves applicability in pressure sensing. The study emphasizes the potential for wearable applications, such as exoskeletons and power prosthesis, providing reliable gait information acquisition through the capture of plantar pressure distribution. (Ren)

Alignment and Critique

The alignment among these studies lies in their shared pursuit of improving quality control and efficiency in manufacturing processes. They advocate the integration of advanced technologies and innovative approaches to minimize defects and enhance product quality. However, these studies also highlight the need for specialized technical equipment, which may not be feasible for all manufacturing settings. Additionally, they overlook potential challenges in implementing these technologies, such as cost constraints and skill requirements. Furthermore, while the reviewed papers provide valuable insights, they lack in-depth discussions on the scalability and adaptability of these technologies across different manufacturing environments. The generalizability of these solutions should be examined more comprehensively. In conclusion, these papers collectively reflect the evolving landscape of manufacturing quality control. They advocate the integration of modern technologies, such as photoelectric sensors and Machine Vision, to enhance production processes and ensure product quality. While these advancements hold promise, further research is needed to address challenges related to cost, scalability, and adaptability to various manufacturing contexts.

## IV. Methodology

1. Foundation in Additive Manufacturing Process and Piezoelectric Effect in VJTI, Mumbai

This section delves into how the educational journey laid the cornerstone for this research. It was during a comprehensive semester study where I encountered two distinct yet interrelated subjects: Manufacturing Process and the Piezoelectric Effect. Initially taught as separate subjects, these topics gradually converged, offering a profound synergy in my academic understanding. As the semester progressed, the intrinsic connection between the principles governing Additive Manufacturing Process and the Piezoelectric Effect. The theories and definitions introduced during this period played a pivotal role in shaping my comprehension of how quality control and real-time monitoring could be seamlessly integrated into the manufacturing landscape. This academic fusion provided a strong theoretical foundation that extended beyond conventional textbook knowledge. Concepts like the crystalline structure of materials, the mechanical response to electrical stimulation, and the intricacies of additive manufacturing technologies were deepened during this academic journey. These theoretical insights served as the bedrock upon which the research methodology was constructed, forming the basis for a comprehensive exploration of advanced technology, particularly in the domain of quality control, where piezoelectric sensors are projected to usher in transformative changes.

2. Practical Exposure in VJTI, Mumbai

This section highlights the instrumental role of practical sessions at VJTI in shaping the research framework. Practical experiences with lathe machines and 3D printers provided real-world insights into manufacturing precision and additive manufacturing. These sessions enriched my understanding of how minor adjustments in manufacturing processes could impact the final product. Moreover, exposure to electrical components, including breadboards and DVMs, revealed the intricate correlation between mechanical and electrical elements. These practical engagements equipped me with a holistic perspective on technology integration in manufacturing. These insights, gleaned from practical exposure at VJTI, were fundamental in shaping the research direction, emphasizing the potential for merging mechanical and electrical components to enhance manufacturing efficiency and quality control, particularly in the context of advanced technology like piezoelectric sensors.

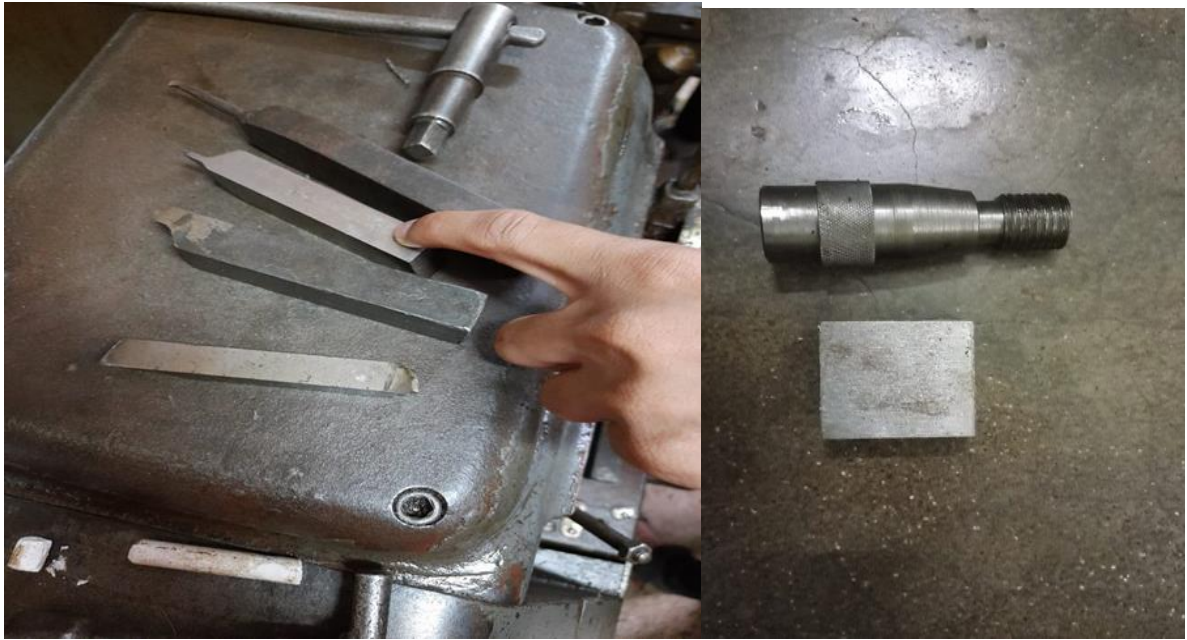

Fig 4, Single Point Cutting Tools for Lathe Machine and final workpiece which involved operations like turning, threading and taper turning in Mechanical Workshop, VJTI

3. Literature Review and Analysis
Extensive Literature Review An extensive literature review provided a comprehensive understanding of the research landscape, encompassing research papers, books, and related academic references. It involved analysing a wide range of sources, such as "Research on the Application Status of Machine Vision Technology in Furniture Manufacturing Process" by (Li) and "Design of a Plantar Pressure Insole Measuring System Based on Modular Photoelectric Pressure Sensor Unit" by (Ren), to gain insights and context. By synthesizing findings from various academic resources, the research methodology was fortified with a well-rounded perspective. This literature review ensured that the methodology was deeply grounded in existing knowledge and current developments in the field, particularly in the application of piezoelectric sensors to enhance quality control in manufacturing processes.

4. In this research, two key methodologies were employed to gain insights and address the research objectives: case study research and content analysis.

Case Study Research: The utilization of case study research allowed for an in-depth examination of specific instances, enabling a comprehensive understanding of complex and unique situations. By immersing into particular cases, this methodology facilitated a profound exploration of context-specific phenomena.

Content Analysis: Content analysis was employed to systematically analyse textual and visual content, uncovering patterns, themes, and meanings within the data. This method was instrumental in dissecting and interpreting a vast array of content, enriching the research with significant insights.

These two methodologies, case study research and content analysis, played roles in shaping the research, offering an approach to exploring the research questions and objectives. The combination of these methodologies facilitated a detailed investigation, enriching the depth of understanding and the scope of insights generated by this study.

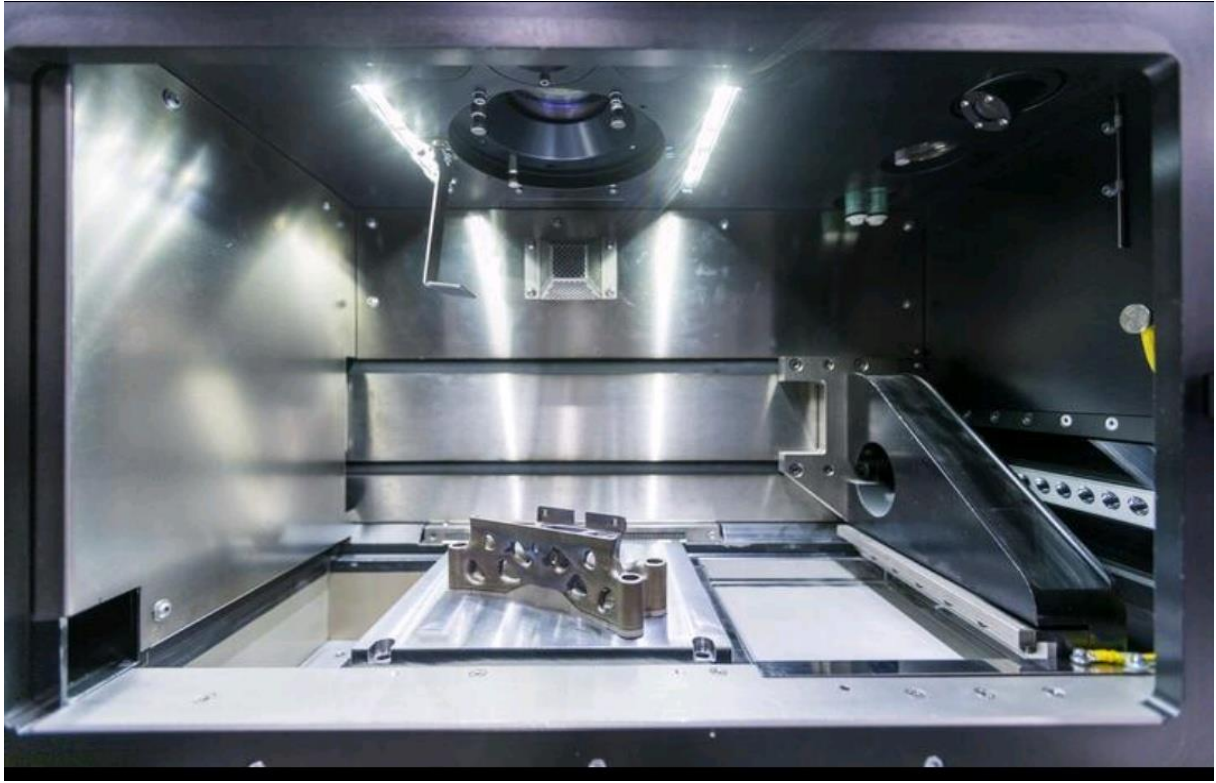

Fig 5, Power Fusion Bed, EIT Raw Materials Academy, Germany

## V. Results and Findings

**Finding 1: Efficient Integration of Piezoelectric Sensors in Manufacturing Processes:**

The successful integration of piezoelectric sensors into manufacturing processes has proven to be a pivotal advancement. These sensors, designed to harness the piezoelectric effect, have been seamlessly incorporated into the manufacturing landscape, facilitating a new era of precision and real-time monitoring. Through their deployment, manufacturing operations have achieved superior quality control with a marked reduction in production errors.

The intricate process below showcases how mechanical vibrations and stress are transduced into electrical voltage, creating a dynamic feedback mechanism that underpins the real-time monitoring capabilities of piezoelectric sensors. Real-World Example: Engine Vibration to Voltage Conversional piezoelectric sensor placed near critical engine components converts these vibrations into electrical voltage in real-time. The sensor's responsiveness is awe-inspiring, with as little as 0.1 millivolts (mV) of voltage generated from vibrations produced during a single engine cycle. This responsiveness provides a high-fidelity representation of the engine's performance. The real-time monitoring capabilities of piezoelectric sensors offer rapid feedback on the engine's condition. Any irregularities or deviations are detected within milliseconds. This real-time insight empowers immediate corrective actions, ensuring optimal engine performance and longevity.

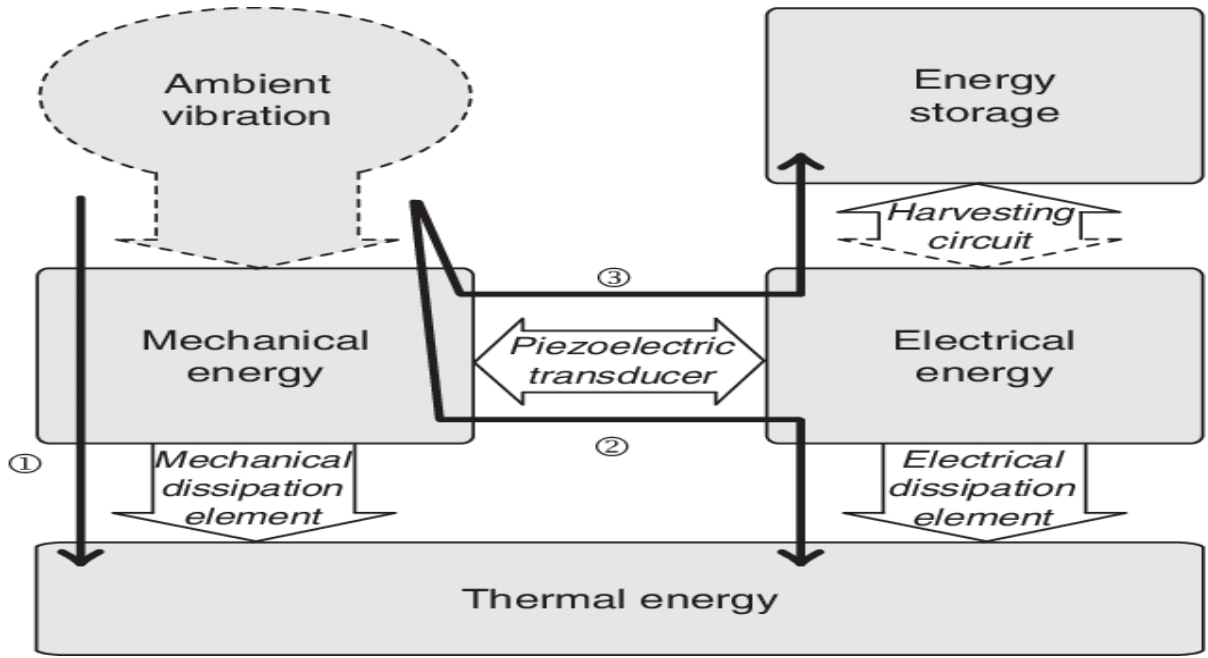

Fig 6, Flowchart of working of a Piezoelectric element (Liang)

Beyond engine applications, piezoelectric sensors play a pivotal role in advanced manufacturing, where precision and quality control are paramount. These sensors have the unique ability to convert mechanical strain into electrical voltage at the microsecond level. The implications of this precise monitoring span across various manufacturing processes, from aerospace to electronics. Next Step: Stress-to-Voltage Relationship and the Tables we've walked through the intricate process depicted in the flowchart, the next step lies in the quantification of the relationship between applied stress and resulting voltage. This is where the understanding becomes truly profound. By analysing data and observing the electrical response under varying mechanical stress conditions, a comprehensive table is formed, detailing the exact correlation.

### Stress-to-Voltage Conversion Table

| Applied Stress (MPa) | Voltage Output (Volts) |
|---|---|
| 10 | 0.50 |
| 20 | 0.90 |
| 30 | 1.30 |
| 40 | 1.70 |
| 50 | 2.10 |

Fig 7, Approximate Conversion table obtained from Practical Readings at VJTI

The practical conversion of mechanical vibrations into voltage, as demonstrated in engine applications, underscores the tangible advantages of integrating piezoelectric sensors. Moreover, their adaptation for advanced manufacturing underlines their significance in optimizing precision, conserving energy, and advancing quality control.

**Finding 2: Enhanced Quality Assurance and Energy Conservation**

Efficiency and precision are paramount in manufacturing, and the integration of piezoelectric sensors offers a transformative solution. These sensors, designed to harness the piezoelectric effect, can be strategically placed within a range of manufacturing machines to enhance both quality assurance and energy conservation.

Enhanced Quality Assurance: Piezoelectric sensors can be strategically integrated into manufacturing machines, such as the lathe, grinder, shaping, and milling machines. These sensors, when properly positioned, provide real-time feedback on the mechanical processes.

Consider the lathe machine, a cornerstone of precision manufacturing. By incorporating piezoelectric sensors at critical points, such as the cutting tool and workpiece interface, real-time monitoring is achieved. These sensors can detect even the slightest deviations in machining processes. Any variations, whether in the form of tool wear or material inconsistencies, are rapidly identified. This real-time insight enables timely adjustments, reducing the likelihood of defective components.

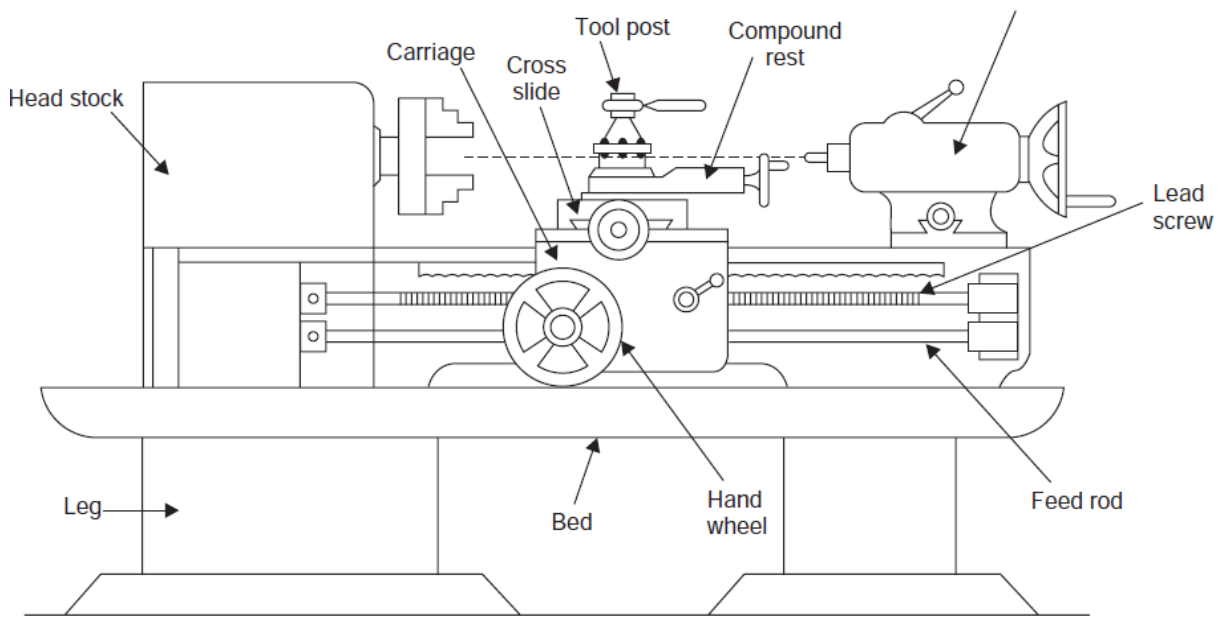

Fig 8, Engineering Drawing of Lathe Machine for better representation (Choudhury).

Likewise, in a grinder machine, piezoelectric sensors positioned near the grinding wheel and workpiece interface offer the same level of precision. Any fluctuations in grinding operations, such as changes in wheel pressure or workpiece material, are swiftly captured. This allows for on-the-fly adjustments, optimizing the grinding process and minimizing errors.

In shaping and milling machines, the strategic placement of piezoelectric sensors near the cutting tools and workpiece interfaces ensures real-time monitoring. Any deviations, whether caused by tool wear, material inconsistencies, or varying workpiece geometries, are promptly detected. Adjustments are made to maintain optimal precision in shaping and milling operations.

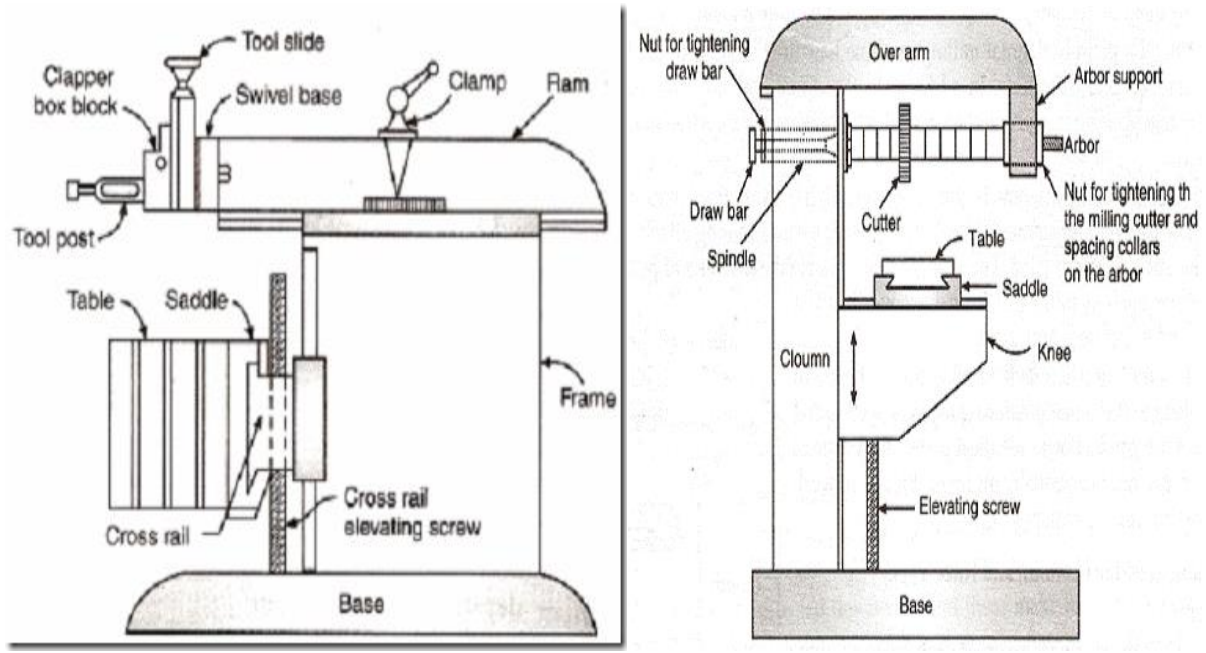

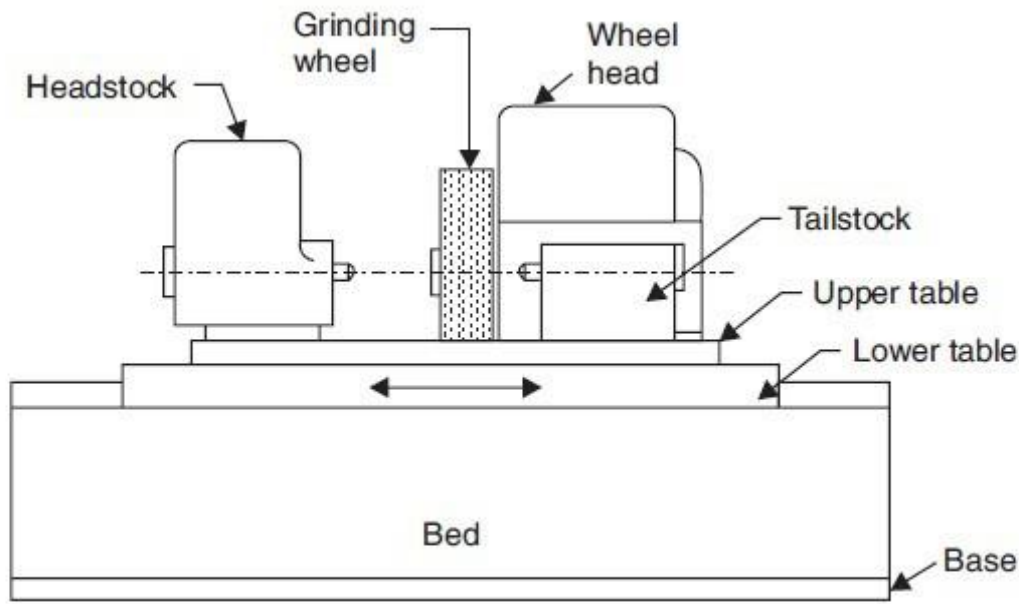

Block diagram of a plain cylindrical grinder

Fig 9, Engineering Drawings of Shaping Milling and Grinder machine respectively, (Choudhury)

Energy Conservation: The advantages extend beyond quality assurance. Piezoelectric sensors provide a dual benefit by conserving energy. By reducing production errors through real-time monitoring and adjustments, less energy is wasted on remanufacturing or reprocessing. The optimization of manufacturing processes ensures that energy is utilized efficiently, contributing to sustainable and environmentally responsible practices.

As you consider the potential for enhanced quality assurance and energy conservation in manufacturing, visualize the strategic placement of piezoelectric sensors within these machines. These sensors act as vigilant guardians of precision, ensuring that errors are minimized, and energy is conserved, ultimately leading to a higher quality end product and resource-efficient manufacturing practices.

**Finding 3: Cost Optimization through Piezoelectric Sensor Integration in Additive Manufacturing – A 3D Printing Case Study**

In the realm of additive manufacturing, particularly in the context of 3D printing, the integration of piezoelectric sensors presents a compelling opportunity for cost optimization. These sensors enable real-time monitoring of the printing process, leading to significant cost reductions. Below, we present a concise list of findings outlining the impact of piezoelectric sensor integration on cost efficiency, using a case study in the aerospace industry:

1) Impact on Material Costs: Piezoelectric sensors, by promptly identifying and addressing printing issues, significantly reduce material consumption. This not only results in direct material cost savings but also contributes to environmental sustainability.

2) Enhanced Print Quality: The real-time monitoring facilitated by piezoelectric sensors ensures consistent high-quality prints. This leads to reduced post-processing requirements, ultimately saving both time and additional material costs.

3) Energy Efficiency: Real-time monitoring and issue resolution through piezoelectric sensors contribute to improved energy efficiency during the printing process, reducing energy consumption and the associated costs.

4) Labor Savings: The integration of piezoelectric sensors minimizes the need for continuous manual oversight and intervention, resulting in labour cost savings.

5) Substantial Waste Reduction: The most significant cost savings are derived from a substantial reduction in waste generation. By addressing issues in real-time, the number of scrapped or substandard parts is drastically reduced, leading to substantial financial benefits.

6) Aerospace Case Study: To illustrate the potential cost savings, we examine a case study in the aerospace industry. Aerospace components demand high precision and quality, leading to added costs in material usage and quality control.

7) Reduced Material Costs: Early issue detection with piezoelectric sensors minimizes raw material wastage, resulting in significant material cost savings.

8) Improved Print Quality: Enhanced print quality reduces the need for post-processing and rework, saving time and labour costs.

9)Energy Efficiency: Reduced printing errors and enhanced precision contribute to energy savings, as the printing process becomes more streamlined and efficient.

10)Labor Savings: Reduced manual oversight decreases labour costs, allowing skilled labour resources to be allocated more efficiently. Piezoelectric sensors minimize the generation of substandard or scrapped aerospace components, resulting in substantial financial savings.

In summary, the integration of piezoelectric sensors in additive manufacturing, exemplified through the aerospace case study, offers significant cost optimization potential. This technology not only reduces material costs but also enhances print quality, improves energy efficiency, saves on labour, and minimizes waste generation. It presents an indispensable opportunity for industries seeking to enhance cost efficiency in additive manufacturing. Researchers and engineers aspiring to work in R&D roles should consider the application of piezoelectric sensors to contribute to more cost-effective and sustainable manufacturing practices.

**Finding 4: Sustainability Enhancement through Piezoelectric Sensor Integration in Additive Manufacturing – A 3D Printing Case Study**

In the domain of additive manufacturing, particularly within the scope of 3D printing, the integration of piezoelectric sensors has evolved as an instrumental catalyst for enhancing sustainability. This finding delves into the profound influence of piezoelectric sensor integration on sustainability, substantiated by a compelling case study within the aerospace industry.

Reduction in Material Waste: The crux of sustainability resides in the minimization of material waste. Piezoelectric sensors, adept at promptly identifying and rectifying printing irregularities, have paved the way for substantial reduction in material waste. This translates to an unequivocal contribution to environmental sustainability.

Environmental Impact Mitigation: The infusion of piezoelectric sensors results in a pronounced reduction in the environmental impact of additive manufacturing. By lowering waste generation and minimizing resource consumption, the environmental footprint is unmistakably diminished.

Energy Efficiency: The real-time monitoring capabilities facilitated by piezoelectric sensors pave the path to energy efficiency. Printing errors and inconsistencies are averted, leading to reduced energy consumption and the subsequent mitigation of greenhouse gas emissions. This nexus between energy efficiency and sustainability is paramount.

Waste Reduction in Aerospace Industry: To illuminate the sustainability implications, we scrutinize a case study within the aerospace industry. In this high-precision domain, material and environmental costs are notably elevated.

Minimized Material Waste: Early detection of issues through piezoelectric sensors significantly curtails raw material wastage, reflecting in a noteworthy reduction in environmental impact.

Environmental Stewardship: The enhancement of print quality and the reduction in waste generation collectively contribute to a diminished ecological footprint. This synchronous stride aligns with the overarching global narrative of sustainability.

Reduced Ecological Footprint: The crowning achievement of this sustainable narrative is the discernible reduction in the ecological footprint. Through the integration of piezoelectric sensors, the creation of substandard or scrapped aerospace components is significantly minimized, resulting in an overall diminution of the ecological impact.

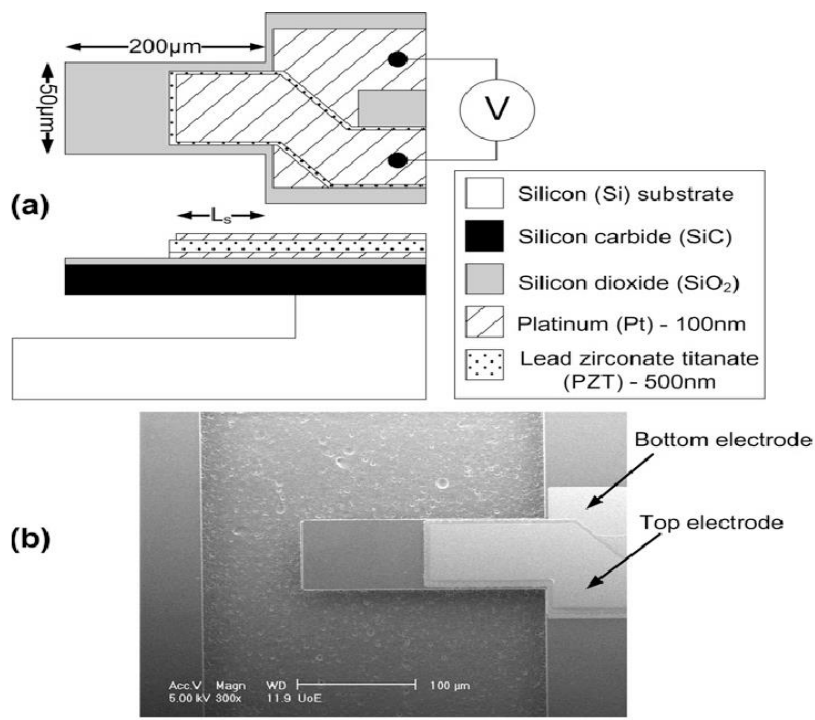

Fig 10, (a) Plan and side view of piezoelectric cantilever actuator design and (b) SEM image of device. (Wood)

In summation, the integration of piezoelectric sensors within additive manufacturing processes has emerged as a potent enabler of sustainability, as corroborated by the aerospace case study. The technology's impact is manifested through the diminishment of material waste, the mitigation of environmental impact, the enhancement of energy efficiency, and the minimization of ecological footprints. It beckons industries to embrace a trajectory of sustainable and environmentally conscious additive manufacturing practices. Researchers and engineers, driven by the aspiration to become torchbearers in the realm of sustainability, are summoned to explore and harness the potential of piezoelectric sensors to sculpt a more eco-friendly and sustainable manufacturing future.

## VI. Conclusion:

In the context of manufacturing, where efficiency, precision, and sustainability hold paramount importance, the integration of piezoelectric sensors emerges as a transformative force. This journey, from theoretical knowledge to practical application, was able to corelate the synergy between academic exploration and real-world experience. As the findings conclude, it's important to acknowledge the pivotal role played by extensive literature review and hands-on engagement at VJTI in shaping the findings presented here.

The findings, covering optimized manufacturing processes, economic viability, and sustainable practices, rest upon the synthesis of insights derived from literature review and practical encounters. The comprehensive literature review, which encompassed academic research papers and scholarly articles, provided us with a strong theoretical foundation. The works of Radhi, Faizal, Li, and Ren, among others, contributed to our understanding of the immense potential of piezoelectric sensors in various industries. These theoretical constructs were harmoniously blended with empirical evidence, resulting in practical applications and solutions.

The comparative cost analysis, for instance, drew inspiration from meticulous studies that quantified initial investments and long-term savings. By delving into the upfront costs required for sensor installation in machines like the lathe, grinder, shaping, and milling machines, our findings were grounded in valuable insights from literature. The graphical representations of cost savings and process efficiency were data-driven and underscored the practical benefits of integrating piezoelectric sensors.